# Quasi-adiabatic Switching for Metal-Island Quantum-dot Cellular Automata

**Géza Tóth and Craig S. Lent**

*Department of Electrical Engineering*
*University of Notre Dame*
*Notre Dame, IN 46556*

## ABSTRACT

Recent experiments have demonstrated a working cell suitable for implementing the Quantum-dot Cellular Automata (QCA) paradigm. These experiments have been performed using metal island clusters. The most promising approach to QCA operation involves quasi-adiabatically switching the cells. This has been analyzed extensively in gated semiconductor cells. Here we present a metal island cell structure that makes quasi-adiabatic switching possible. We show how this permits quasi-adiabatic clocking, and enables a pipelined architecture.

# I. Introduction

In recent years the development of integrated circuits has been essentially based on scaling down, that is, increasing the element density on the wafer. Scaling down of CMOS circuits, however, has its limits. Above a certain element density various physical phenomena, including quantum effects, conspire to make transistor operation difficult if not impossible. If a new technology is to be created for devices of nanometer scale, new design principles are necessary. One promising approach is to move to a transistor-less cellular architecture based on interacting quantum dots, Quantum-dot Cellular Automata (QCA, [1-5]).

The QCA paradigm arose in the context of semiconductor quantum dots, usually formed by using metallic gates to further confine a two-dimensional electron gas in a heterostructure. The quantum dots so formed exhibit quantum confinement effects and well separated single-particle eigenstates. The QCA cell consists of four (or five) such dots arranged in a square pattern. The semiconductor implementation has significant advantages in that both the geometry of the dots and the barrier-heights between the dots can be tuned by adjusting gate potentials. QCA switching involves electrons tunneling through interdot barriers to reconfigure charge in the cell. Information is encoded in the arrangement of charge within the cell.

Quite early in the development of QCA ideas it was realized that the quantization of energy levels in the dots, is not crucial to QCA operation. All that is really required is (approximate) charge quantization on the dot, and quantum-mechanical tunneling to enable switching. The robustness of the QCA scheme is due in large measure to the fact that the information is contained in classical degrees of freedom, while quantum effects simply provide the "grease" that enables switching to occur. It was shown theoretically that in principle, metallic islands connected by capacitive tunnel junctions could also be used to realize QCA cells [6].

The semiconductor QCA implementation has remained the focus of development as the theory has been extended to large arrays of devices and computer architecture questions. A key advance was the realization that by periodically modulating the inter-dot barriers, clocked control of QCA circuitry could be accomplished. The modulation could be done at a rate which is slow compared to inter-dot tunneling times, thereby keeping the switching cells very near the instantaneous ground state. This quasi-adiabatic switching [3] paradigm has proven very fruitful. Quasi-adiabatic clocking permits both logic and addressable memory to be realized within the QCA framework. It allows a pipe-lining of computational operations.

Recently, the first experimental realization of a functioning QCA cell has been reported [7]. This was accomplished in the metal-dot system. The bistable behavior and full cell operation were confirmed. This experimental success raises the question as to whether the quasi-adiabatic switching can be implemented in the metal-dot system. The barriers between dots in this system are typically very thin slices of oxide. While there have been some promising experiments involving the modulation of such barriers[8], in general it is much harder to accomplish than in the semiconductor case. In this paper we demonstrate a scheme for quasi-adiabatic switching of metallic QCA cells. The modulated barrier is basically replaced by another dot, whose potential can be altered.

Section II includes a brief description of the semiconductor QCA, adiabatic switching and the metal-island QCA cell. This section is necessarily brief; a fuller review is given in Ref. [3]. In Sec. III we show how adiabatic switching can be realized with a newly designed metal island cell. The theoretical model of metal island circuits will be discussed in Sec. IV as well as applications of the model to pipe-lined quasi-adiabatic shift-registers.

# II. Quantum-dot Cellular Automata

## A. Semiconductor QCA

The semiconductor QCA cell consists of four quantum dots as shown in Fig. 1(a). Tunneling is possible between the neighboring dots as denoted by lines in the picture. Due to Coulombic repulsion the two electrons occupy antipodal sites as shown in Fig. 1(b). These two states correspond to polarization +1 and -1, respectively, with intermediate polarization interpolating between the two.

In Fig. 1(c) a two cell arrangement is shown to illustrate the cell-to-cell interaction. Cell 1 is a driver cell whose polarization takes the range -1 to 1. It is also shown, how the polarization of cell 2 changes for different values of the driver cell polarization. It can be seen, that even if the polarization of the driver cell 1 is changing gradually from -1 to +1, the polarization of cell 2 changes abruptly from -1 to +1. This *nonlinearity* is also present in digital circuits where it helps to correct deviations in signal level: even if the input of a logical gate is slightly out of the range of valid "0" and "1" voltage levels, the output will be correct. In the case of the QCA cells it causes that cell 2 will be saturated (with polarization close to -1 or +1) even if cell 1 was far from saturation.

A one-dimensional array of cells[4] can be used to transfer the polarization of the driver at one end of the cell line to the other end of the line. Thus the cell line plays the role of the wire in QCA circuits. Moreover, any logical gates (majority gate, AND, OR) can also be implemented, and using these as basic building elements, any logical circuit can be realized[5].

## B. Adiabatic switching with semiconductor QCA

In this paradigm of ground state computing, the solution of the problem has been mapped onto the ground state of the array. However, if the inputs are switched *abruptly*, it is not guaranteed that the QCA array really settles in the ground state, i.e., in the global energy minimum state. It is also possible, that eventually it settles in a *metastable* state because the trajectory followed by the array during the resulting transient is not well controlled.

This problem can be solved by adiabatic switching [3] of the QCA array, as shown schematically in Fig. 2. Adiabatic switching has the following steps: (1) before applying the new input, the height of the interdot barriers is lowered thus the cells have no more two distinct polarization states, P=+1 and P=-1. (2) Then the new input can be given to the array. (3) While raising the barrier height, the QCA array will settle in its new ground state.

The quasi-adiabaticity of the switching means that the system is very close to its ground state during the whole switching process. It does not reach an excited state after setting the new input, as happens if the input is simply switched abruptly. Since the system does not get to an excited state from the ground state the dissipation to the environment is minimal. On the other hand, to maintain quasi-adiabaticity the time over which the barrier height is modulated must be long compared to the tunneling time through the barrier. Typically a factor of 10 reduces the non-adiabatic dissipation to very small levels.

The previous structure can be used for processing a series of data, as shown in Fig.3. While changing the input, the barrier is low therefore the cells do not have a definite polarization. Then the barrier height is increasing, until it reaches the value, where the cell polarization is fixed. This means that the barriers are so high that the interdot tunneling is not possible, the polarization of the cells keeps its value independent of the effects of the external electrostatic fields. At that point the output can be read out. Then the barriers are lowered again, and the next input can be given to the array. Fig. 3(c) shows the input and output flow for this case.

The cells of such a QCA array have three *operational modes*: if the barriers are low then the cell does not have distinct polarization. This can be called the *null operational mode*. If the barriers are high then the polarization of the cells does not change. This can be called the *locked operational mode*. In case of intermediate barrier heights, the *active mode*, the cells have two distinct polarization states: P=+1 and P=-1, however, external electrostatic field (due to the effects of the neighboring cells) can switch it from one polarization to the other. The operational modes are summarized in Fig. 4. Thus the cells periodically go through the null→active→ locked→active→null series.

The arrangements shown in Fig. 3 can be expanded for more QCA sub-arrays working in a *pipeline* structure as shown in Fig. 5. Now each sub-array reads the output of the left neighbor when the neighbor is in locked state and begins to write into their right neighbor when it is in null state. The main advantage of the pipeline architecture is that the computations with the new input start before the computations with the old input are finished. Each unit gives its subresult to the following unit and then begins to process the subresult of the previous unit.

The barrier heights of the arrays are controlled by four different clock signals. The clock signal given to an array is delayed by 1/4 period time relative to the clock signal of its left neighbor.

With only these four clock signals, even more sophisticated pipeline structures containing logical gates and flip-flops can be realized [3].

## C. Metal-island QCA

QCA cells can be also built from metallic tunnel junctions and very small capacitors[6]. There are two main differences between the semiconductor and the metal dot QCA's. (1) Capacitively coupled metal islands are used rather than Coulombically coupled quantum-dots. Unlike the quantum dot, the metal island contains many conduction band electrons. (2) A classical capacitive model can be applied instead of a Schrödinger-equation model.

The only non-classical phenomenon is the tunneling of electrons between metal islands through tunnel junctions. The metal islands have a special feature: they are connected to the other islands through tunnel junctions. If these tunnel junctions were replaced by capacitors the island charge would be zero; however, through the tunnel junction an integer number of electrons can tunnel into or out of the island. Thus the charge of an island is integer multiple of the elementary charge.

In the case of the metal island cell it is helpful to first consider first a double-dot, two of the islands as the basic building element rather than a whole cell of four dots. The two metal islands ("dots") connected by a tunnel junction give a *bistable circuit element* (See Fig. 6(a), framed double dot). Depending on the input voltages, the excess electron will show up either at the upper dot or at the lower dot. By setting the input voltages the occupancy of these dots can be determined, that is, we can set the "polarization" of this bistable element. (Let +1 and -1 denote the two possible polarizations.)

As shown in Fig. 6(a), a *QCA cell* consists of two of these bistable elements or *half cells*. It can have two polarizations: +1 if the two excess electrons are in the upper right and lower left islands, -1 if they are in the other two islands (Fig. 6(b)). If several of these cells are placed in a line and they are coupled capacitively then by switching the input voltage of the first cell a polarization change will be transmitted along the cell line as in the case of the semiconductor cell. All logical and computational structures which can be implemented with the semiconductor QCA can also be realized with the metal island cells.

# III. Quasi-adiabatic switching with metal-island cells

The circuit for the metallic half-cell is shown in Fig. 7(a). It contains three metallic islands. The occupancy of the three islands is represented by a triple of integers [n1 n2 n3]. During operation its occupancy can be [100], [010] or [001], as shown in Fig. 7(b). The [100] charge configuration corresponds to the polarization +1 case, the [001] charge configuration corresponds to the polarization -1 case, while [010] represents a null polarization.

The top and bottom islands are biased with respect to ground through (non-leaky) capacitors. The bias voltage raises the electrostatic potential of these islands (lowering electron potential energy) so that an electron is attracted from ground into the three-island chain. The top and bottom islands can be viewed as a double well system with the middle island acting as a controllable barrier.

Each of the three islands has a corresponding gate electrode. A differential input is applied to the gate electrodes for the top and the bottom islands. The half cell can be switched from one polarization state to the other by this input voltage. The input can be supplied externally or from another half-cell (as discussed in the next section). The voltage on the gate electrode for the middle island is used as a control. The three operational modes of the half cell (active, locked and null) can be selected by setting this voltage to one of three discrete levels corresponding to the three modes.

The three operational modes are shown schematically in Fig. 8(a-c). The switching in active mode is illustrated in Fig. 8(a). First the pictorial representation of the process can be seen, then the energies of the [100], [010] and [001] configurations are given during the switching. The differential input bias changes from positive to negative. Initially, the top electrode is at a positive potential while the bottom electrode is negative resulting in the [100] configuration having the lowest energy. As V decreases, the energy of the [100] configuration increases and will be higher than that of the [010] configuration. Thus the electron tunnels from the top island to the middle island, and the three-island system is in the [010] configuration. Decreasing V further, the [001] will be the minimal energy configuration, and therefore the electron tunnels to the bottom island. In Fig. 8(b) the locked operational mode is illustrated. The control electrode has a lower potential (higher electron potential energy) than in active mode, so the electron cannot get to the middle island from the top one. In null mode the control electrode is at a higher potential (lower electron potential energy) than in active mode thus the electron stays in the middle island regardless of the applied differential input bias as shown in Fig. 8(c). [9]

# IV.  Theory of operation

## A. Physical model

We can model the quasi-static behavior of the circuits described by considering only the energy of the various charge configurations of the system. We treat here only the zero temperature situation. The system is composed of gate electrodes and metal islands, coupled by tunnel junctions and capacitors[10,11]. The gate electrode voltages are fixed by external sources, and the charge on each metal island is constrained to be an integral multiple of the fundamental charge. The electrostatic energy of a configuration can be expressed in terms of the voltages and charges on gate electrodes and metal islands.

$$E \; = \; \frac{1}{2} \begin{bmatrix} q \\ q' \end{bmatrix}^T C^{-1} \begin{bmatrix} q \\ q' \end{bmatrix} - v^T q' \qquad (1)$$

Here $C$ is the capacitance matrix for the islands and electrodes, $v$ is a column vector of voltages on the gate electrodes, $q$ and $q'$ are the column vectors of the island charges and the lead charges, respectively. The first term of the energy expression describes the electrostatic energy stored in the capacitors and tunnel junctions. The second term is the work done by the sources transferring charge to the leads. The equilibrium charge configuration for T=0 K temperature minimizes this electrostatic energy.

For a QCA cell to be switched quasi-adiabatically, input and clock voltages are varied smoothly enough so that the cell is very close to its equilibrium ground state configuration during the time it is switching. Thus during the *active* mode of cell operation, the cell should be in the configuration which minimizes the total electrostatic energy for the cell. The same is true for the *null* mode.

The *locked* mode, by contrast, is designed to provide a short-term memory, *i.e.*, the cell configuration is held to what it was in the immediate past so that the locked cell can be used as a fixed input for another cell which is being switched. Thus it is by design not necessarily in the minimum energy configuration but may be in a metastable state. To model this requires knowing not just the minimum energy configuration, but also the allowed transitions between various configurations. For the QCA half-cell, the six basic *allowed transitions* are summarized in Fig. 9. Notice that there is no transition directly from the top island to the bottom island. This is important for the operation of the locked mode. Suppression of this transition is the reason that there is no direct tunneling path between either the top or bottom electrode and ground.

We can treat all these modes using a single modeling algorithm. As the input voltages are changed in small steps, at each step we examine whether an allowed transition could decrease the energy of the system. If so then the tunneling event takes place instantaneously and the dot occupancies change. This approach is only applicable to the quasi-adiabatic situation we consider here. Refinements which would extend these calculations to high-frequencies would include specific tunneling rates in a Monte Carlo [12,13] or master-equation[14] approach and would include co-tunnelling[15,16] rates.

## B. Operational modes

For simulations shown below parameters for capacitors and voltage sources were chosen in the range of practically realizable values for metal islands fabricated with Dolan shadow-evaporation techniques. They are also chosen in the design space to fulfill the requirement for a reasonable range for the input and control voltages. We have performed numerical simulations of the switching of a half-cell using the model described above. The specific parameter values used were: $C=420aF$, $C_1=300aF$, $C_2=25aF$, $C_3=80aF$, $C_4=200aF$ and $U=0.36mV$. With this set of parameters the control voltages corresponding to locked, active and null operational modes are $V_c = -0.18$, $0.18$ and $0.68mV$, respectively. The input bias changes in the range of $-0.3$ and $+0.3mV$.

In Fig. 10 the transfer characteristics of this half cell can be seen in active mode, that is, for $V_c$=0.18mV. It is piecewise linear, and the abrupt change in value and slope is due to tunneling events, thus the nonlinearity of the transfer characteristics comes from the charge quantization on the metal island.

It is instructive to construct a diagram of the system state as a function of the input voltage and $V_c$. Fig. 11 shows the equilibrium ground state "phase diagram" for the system as a function of these two voltages. For the null and active mode, this is sufficient information to characterize the switching behavior. However for the locked mode, we must assume a particular starting point. Fig. 12 shows this state diagram for the case when the input voltage is increasing from -0.45 mV to +0.45mV. For $V_c$ chosen to keep the system in the locked mode, this means that the system is initially in the [001] state and is kept there. The opposite situation is depicted in Fig. 13, where the system starts with a positive input voltage and is thus in the [100] case. The locked mode keeps it there because the [100]→ [001] transition is suppressed.

The three operational modes will be analyzed using the state diagram shown in Fig 12. Taking $V_c$=0.18mV the circuit is in active operational mode. Following the arrow belonging to the 0.18mV level, the change of the charge configurations as V changes from +0.3 to -0.3mV can be read from the graph. The transition series belonging to this case is [001]→[010]→[100]. The electron tunnels from the third island to the second island, and then moves further to the first island.

If $V_c$ is decreased to -0.18mV, the potential of the middle electrode also decreases and the electron from the islands on the sides can not get to the middle island. This is the locked operational mode, the occupancy does not change even if the V bias voltage is changed, as can be seen following the bottom arrow in Fig. 12.

If $V_c$ is increased from the value it had in case of active mode to 0.68mV, then the electron will be drawn to the middle island. It will stay there independent of the input voltages, as can be seen if one follows the top arrow in Fig. 12. This is the null operational mode.

The critical points on these state diagrams are labeled T and M. The values of V and $V_c$ for these points can be given analytically in terms of the circuit parameters. If we let $X = C + C_2 + C_4$ and $Y = 2C + C_2 + C_3$, then

$$V^T = 0, \tag{2}$$

$$V_c{}^T = \frac{e}{2C_3}\left(\frac{\dfrac{Y}{C}\left(2\dfrac{UC_4}{e}-1\right)-4\dfrac{UC_4}{e}+\dfrac{X}{C}+\dfrac{C}{X}}{\dfrac{X}{C}-1}\right), \tag{3}$$

$$V^M = \frac{e}{2C_2}\left(\frac{\left(\dfrac{X}{C}-1\right)^2}{\dfrac{XY}{C^2}-2}\right), \tag{4}$$

$$V_c{}^M = \frac{e}{2C_3}\left(\frac{\dfrac{Y}{C}\left(2\dfrac{UC_4}{e}-1\right)-4\dfrac{UC_4}{e}+2}{\dfrac{X}{C}-1}\right) \tag{5}$$

It is worthwhile to note that for higher $V_c$ values similar graph to Fig. 12 could be drawn, except for that the [100], [010] and [001] phases would be replaced by the [110], [020] and [011] phases, respectively. If $V_c$ is increased further, then the [120], [030], phases [021] can be found in the diagram. Thus the only difference in the system behavior for higher (lower) $V_c$ values is that the population of the middle island is increased (decreased) by a constant. In this way it can be said that the system behavior is periodic in $V_c$, and it is not more informative to draw a graph for a wider range of control voltages. The $\Delta V_c$ periodicity of the phase diagram in the $V_c$ direction is:

$$\Delta V_c = \frac{e}{C_3} \tag{6}$$

## C. QCA shift register

We construct a simulation of a cell line acting as a shift register, that is a 1D array of capacitively coupled QCA cells. A QCA cell consists of two half cells as depicted in Fig. 14(a). It can have three different occupancies: [001 100] for P=+1 polarization, [100 001] for P=-1 polarization and [010 010] for the null state as shown in Fig. 14(b). The adiabatic switching is realized with four different clock signals as it is shown in Fig. 15. Due to these clock signals the operational mode of a half-cell in the line changes periodically: active→locked→active→null.

The operation of a line of four cells can be seen in Fig. 16. Each line of the graph shows the polarization of a cell as a function of time. In the figure the parts are framed where the cells are in locked operational mode. The state of the cell can be considered valid only in this state, that is, it is supposed to be read externally only during this time.

The shift register is instructive because in principle each element could be replaced by a more complex computational unit. This is how more sophisticated processing could be achieved in this paradigm. The designs of larger-scale functional units as reviewed in Reference [3] can now just be taken over with this new cell design.

# V. Conclusions

In this paper a structure was proposed to realize the adiabatic switching with metal-island QCA cells. Adiabatic switching provides a solution for the crucial problem of ground state computing, namely, that a larger system may settle in a metastable state instead of the ground state. It also makes pipelining and constructing large, digital-like QCA circuits possible.

The core of the proposed QCA cell is a bistable element consisting of three metal islands, tunnels junctions, and capacitors. Its operation was presented in a simulation example, on the basis of phase diagrams. Beside an individual half cell the operation of a cell line was also shown.

## Acknowledgments

We gratefully acknowledge stimulating conversations with members of the Notre Dame Nanoelectronic Group, especially John Timler.

**Figure Captions**

FIGURE 1. Schematic of the basic four-site semiconductor QCA cell. (a) The geometry of the cell. The tunneling energy between two sites (quantum dots) is determined by the heights of the potential barrier between them. (b) Coulombic repulsion causes the two electrons to occupy antipodal sites within the cell. These two bistable states result in cell polarization of P=+1 and P=-1. (c) Nonlinear cell-to-cell response function of the basic four-site cells. Cell 1 is a driver cell with fixed charge density. In equilibrium the polarization of cell 2 is determined by the polarization of cell 1. The plot shows the polarization $P_2$ induced in cell 2 by the polarization of its neighbor, $P_1$. The solid line corresponds to antiparallel spins, and the dotted line to parallel spins. The two are nearly degenerate especially for significantly large values of $P_1$.

FIGURE 2. The steps of the quasi-adiabatic switching are the following: (1) before applying the new input, the height of the interdot barriers are lowered thus the cell have no more two distinct polarization states, P=+1 and P=-1. (2) Then the new input can be given to the array. (3) While raising the barrier height, the QCA array will settle in its new ground state. The adiabaticity of the switching means that the system is very close to its ground state during the whole process. It does not get to excited state after setting the new input, as it happened in the case of non-adiabatic switching. Since the system does not get to an excited state from the ground state the dissipation decreased a lot.

FIGURE 3. QCA structure for the processing of data series. (a) The schematic of the structure. (b) The clock signal given to the cells to control their interdot barrier height. (c) The input and output data flow. The new input is given to the array when the barriers are low and the output is read out of the array when the barriers are high, and the polarization of the cells is fixed. (H, M and L stand for 'high', 'medium' and 'low', respectively.)

FIGURE 4. The three operational modes of the QCA cell in case of adiabatic switching. In active mode, the cells have two distinct polarizations: P=+1 and P=-1, and the external electrostatic filed can switch cells from one polarization to the other. In locked mode, the interdot barriers are high therefore the polarization of the cell cannot be switched, it is fixed. In null mode, the barriers are low thus the cell does not have a definite polarization.

FIGURE 5. Pipeline architecture with QCA arrays. (a) All of the arrays get the input from the left neighbor and give the output to the right neighbor. (b) The clock signals used for the control of the interdot barrier height. Each array gets the clock signal delayed by 1/4 period time relative to its left neighbor. Even more sophisticated structures containing logical gates and flip-flops need no more than four different clock signals.

FIGURE 6. Metal-island QCA cell. (a) The QCA cell consists of two capacitively coupled bistable elements. Such a bistable element consists of two metal islands. The excess electron can be either in the top or in the bottom island, giving the two possible charge configurations. (b) Symbolic representation of the two possible polarizations of the QCA cell.

FIGURE 7. Metal island half-QCA structure permitting adiabatic switching. (a) The circuit consists of three metal islands connected to each other by tunnel junctions. Each island has a capacitively coupled electrode. Applying the V differential input bias and the $V_c$ control voltage the occupancy of the dots can be determined. The middle island is grounded in order to provide an excess electron in the three island system that is necessary to realize the [100]/[010]/[001] charge configurations. The two voltage sources are used to increase the potential of the top and bottom islands to make the switching more abrupt. (b) The symbolic representation for the three island system. The occupancies corresponding to the P=+1, P=-1 polarizations and the null state (indefinite polarization) are shown.

FIGURE 8. The three operational modes. (a) Active operational mode. The electron tunnels from the top island to the bottom island through the middle island, if electrode voltages change. First the pictorial representation of this process is shown. '+', '-' and '0' refers to the sign of the electrode voltages. Then the energies of the [100], [010] and [001] charge configurations can be seen during the switching. The dot refers to the charge configuration the system occupies. (b) Locked operational mode. The electron is locked in either the top or the bottom island, because the [010] configuration has much higher energy than the others. (c) Null operational mode. The electron is locked in the middle island, because the [010] configuration has much lower energy then the other two.

FIGURE 9. The six basic tunneling events that can happen in the three-island structure shown in Fig. 7

FIGURE 10. Transfer characteristics of the half cell structure given in Fig. 7 for active mode. It is piecewise linear, and the abrupt change in value and slope are due to tunneling events. In case of a metal island QCA the nonlinearity comes from the charge quantization. Replacing the tunnel junctions with linear capacitors the circuit would also be linear.

FIGURE 11. Phase diagram of the half cell structure permitting adiabatic switching. The minimal energy configuration is shown as a function of the differential input bias V and the control voltage $V_c$. The control voltage level of the locked operational mode is also shown.

FIGURE 12. State diagram of the half cell for switching. It shows the occupancies of the three islands as the V differential input bias *increased* from -0.45 to +0.45mV for a $V_c$ range of -0.3 and 0.7mV. The voltage levels for the three operational modes are also shown. The charge configuration is also given for each region of the diagram. Note, that the [001] and the [100] phases seem to have a common border, but there is a very "thin line" of [010] or [1,-1,1] phase between them. (The direct transition from [001] to [100] is not possible.) The dots with a "+" sign refer to -1 electron on the dot that is an excess positive charge.

FIGURE 13. State diagram of the half cell for switching. The diagram shows the occupancies of the three islands as the V differential input bias *decreased* from +0.45 to -0.45mV for a $V_c$ range of -0.3 and 0.7mV. The voltage levels for the three operational modes are also shown. The charge configuration is also given for each region of the diagram. Comparing with Fig. 12, the differences are due that V changes in the opposite direction.

FIGURE 14. Metal-dot QCA cell. (a) It consists of two half cells that get the same control voltage. (b) The occupancies corresponding to the P=+1, P=-1 polarizations and the null state.

FIGURE 15. QCA cell line with quasi-adiabatic switching. (a) The four different clock signals applied to the control input of the cells. They are shifted relative to each other by 1/4 period time. The $V_c$=-0.18 and 0.68mV levels correspond to the locked and null operational mode, respectively. (b) The schematic of the QCA cell line. The control voltages applied to the half cells are shown for each of them.

FIGURE 16. Simulation of a QCA line of four cells. The top plot shows the V differential input bias of the first cell as the function of time, the other four

graphs are the polarizations of the cells. ("0" refers to the null state.) The polarization of a cell is valid if it is in the locked operational mode. In this case the polarization is shown in the frame in the graph. Each cell follow their left neighbor's polarization with a delay.

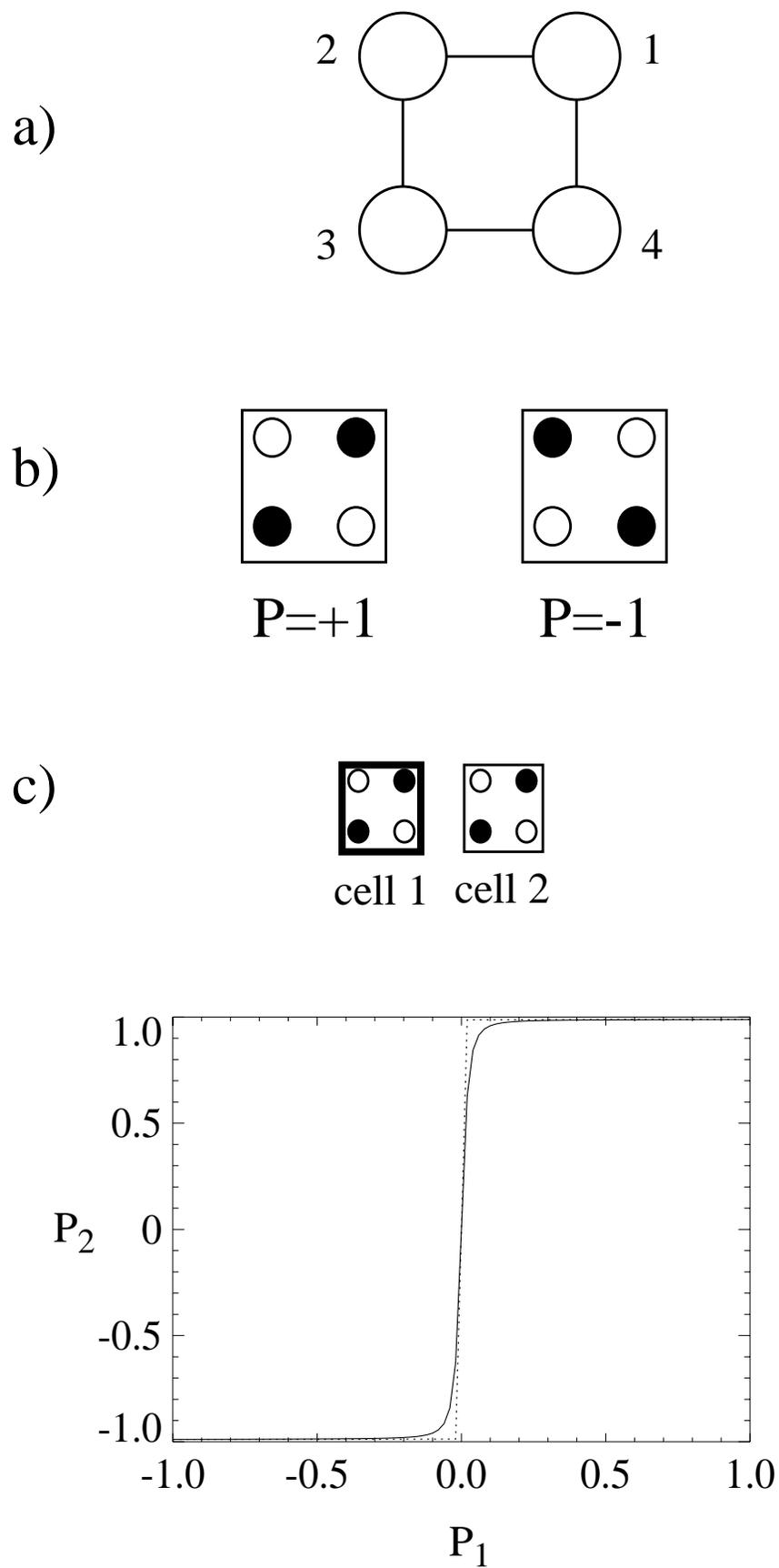

**Remove old input**

**Applying new input**

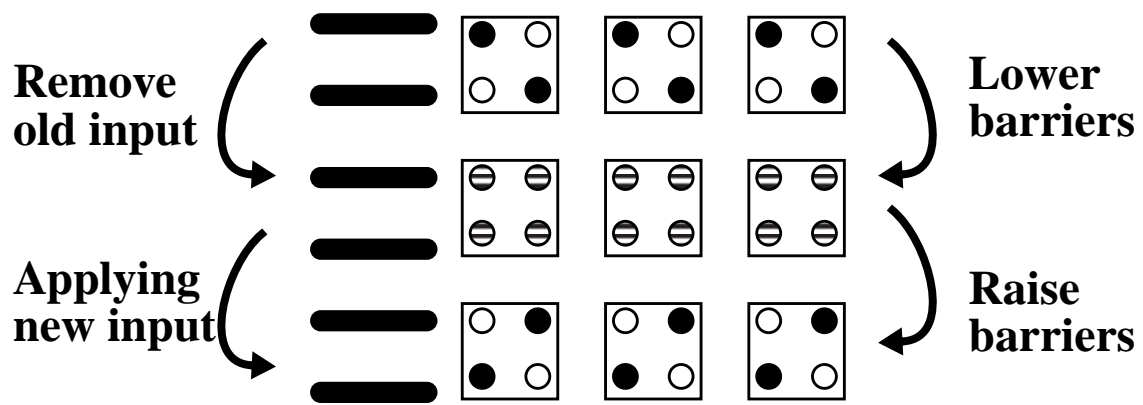

**Lower barriers**

**Raise barriers**

Figure 2

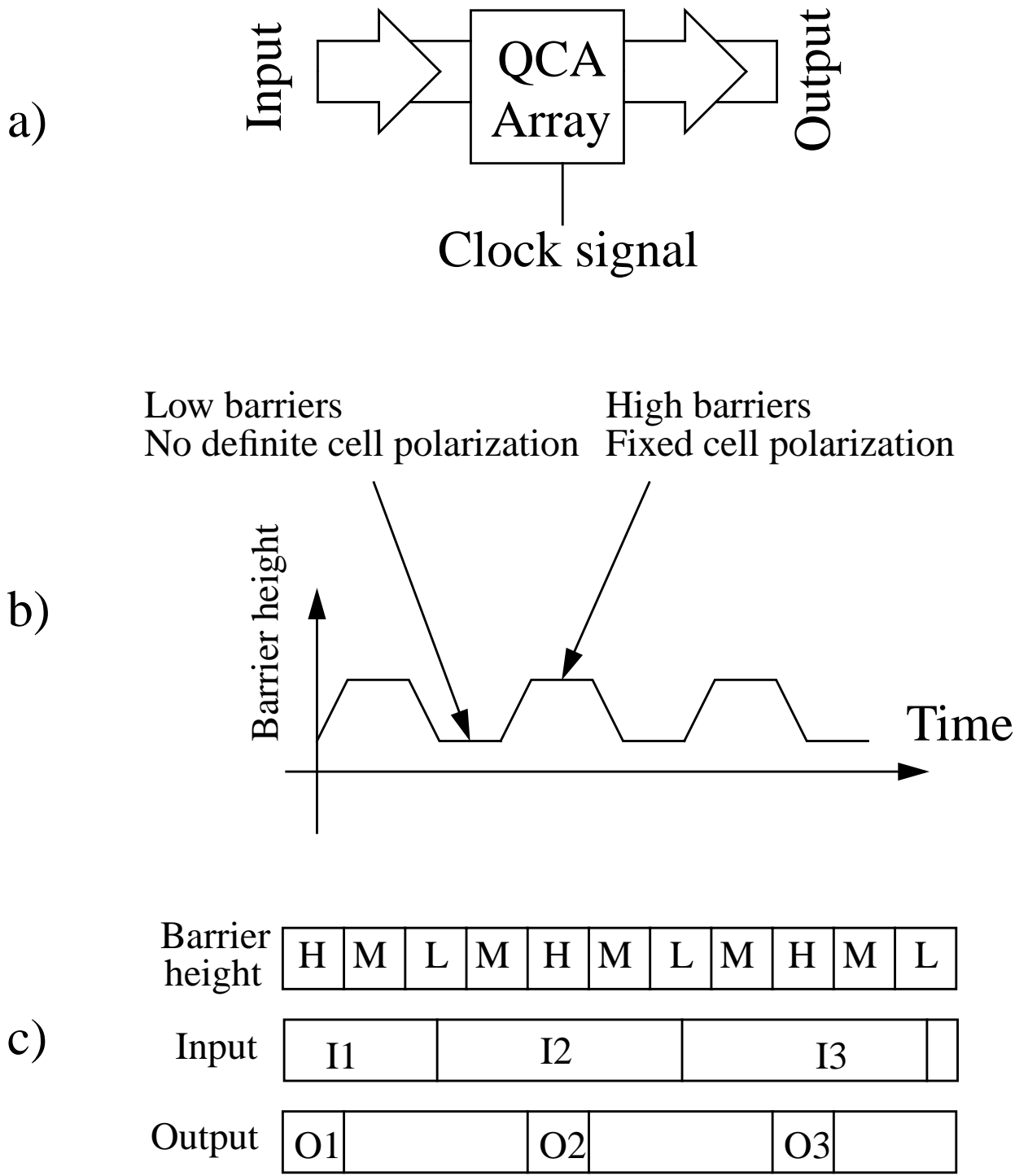

a)

Input → QCA Array → Output

Clock signal

b)

Low barriers
No definite cell polarization

High barriers
Fixed cell polarization

Barrier height

Time

c)

| Barrier height | H | M | L | M | H | M | L | M | H | M | L |
|---|---|---|---|---|---|---|---|---|---|---|---|
| Input | I1 | | | I2 | | | | I3 | | | |
| Output | O1 | | | O2 | | | O3 | | | | |

Figure 3

| Operational mode | Barrier height | Cell polarization |
|---|---|---|
| Active | Medium | Between +1 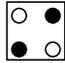 and -1 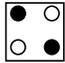 |
| Locked | High | +1 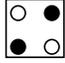 or -1 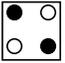 |
| Null | Low | Indefinite 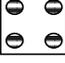 |

Figure 4

a)

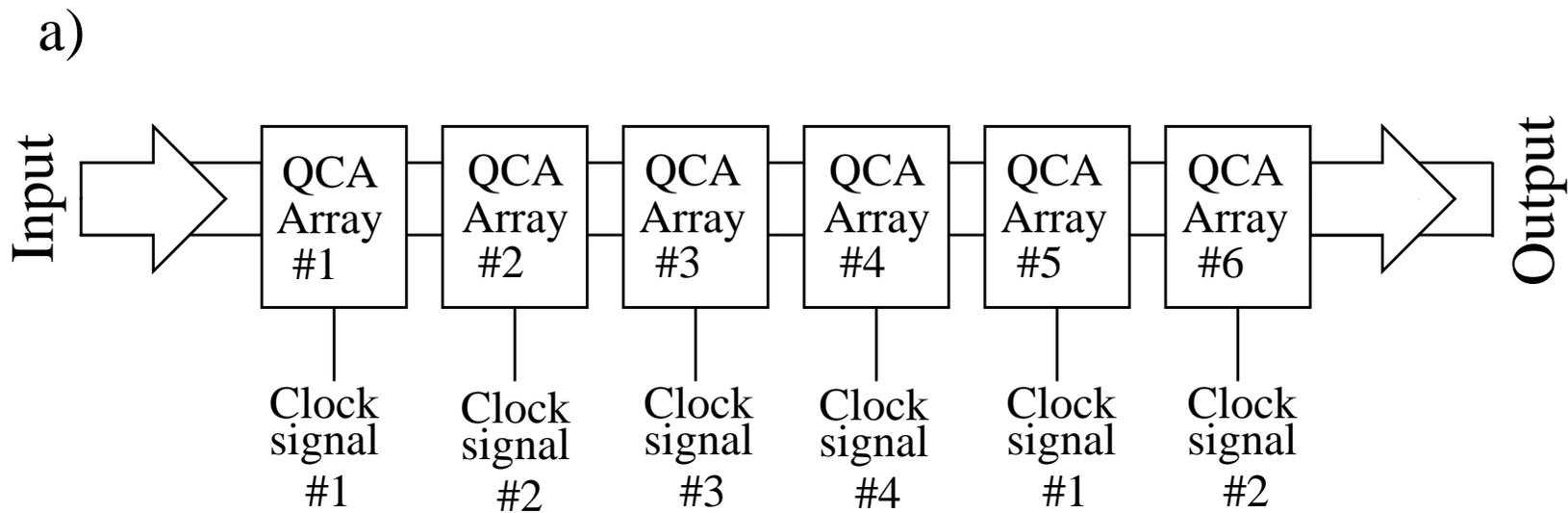

b)

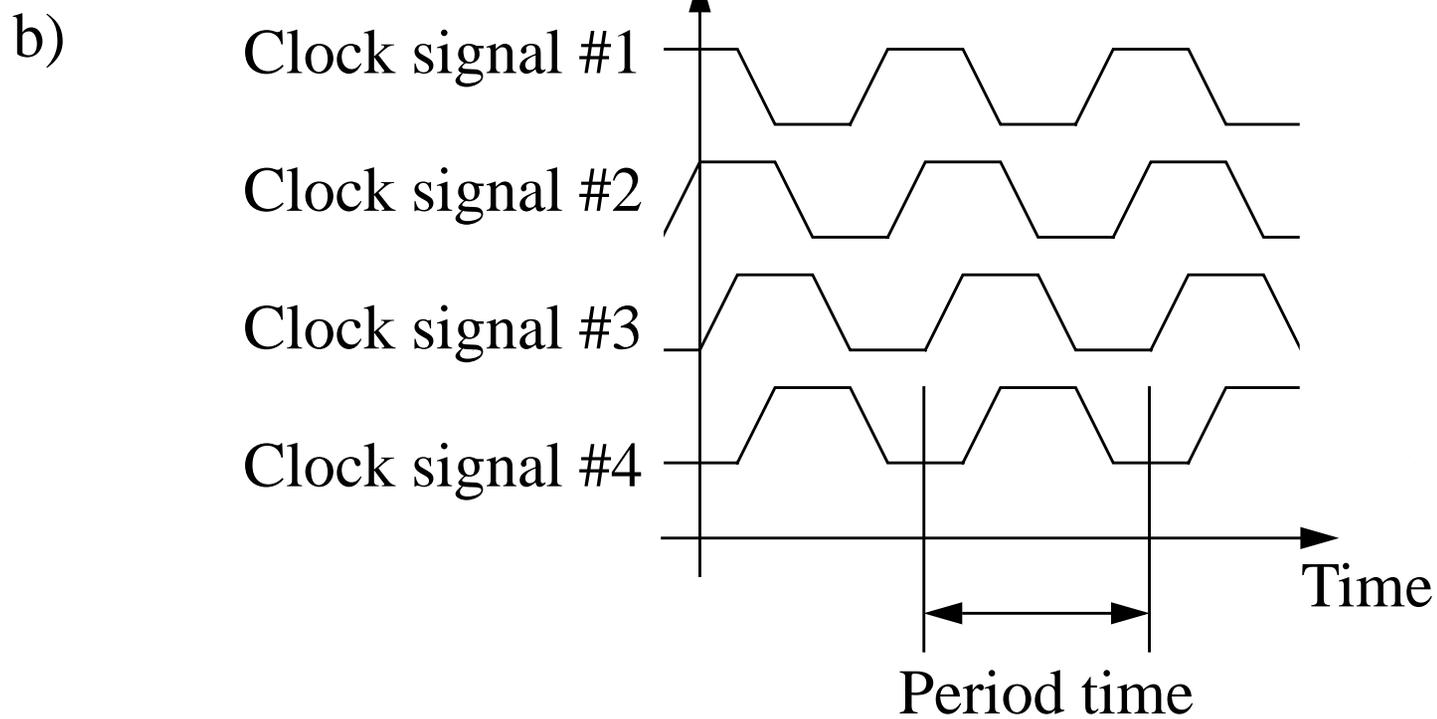

Figure 5

Two-island bistable element

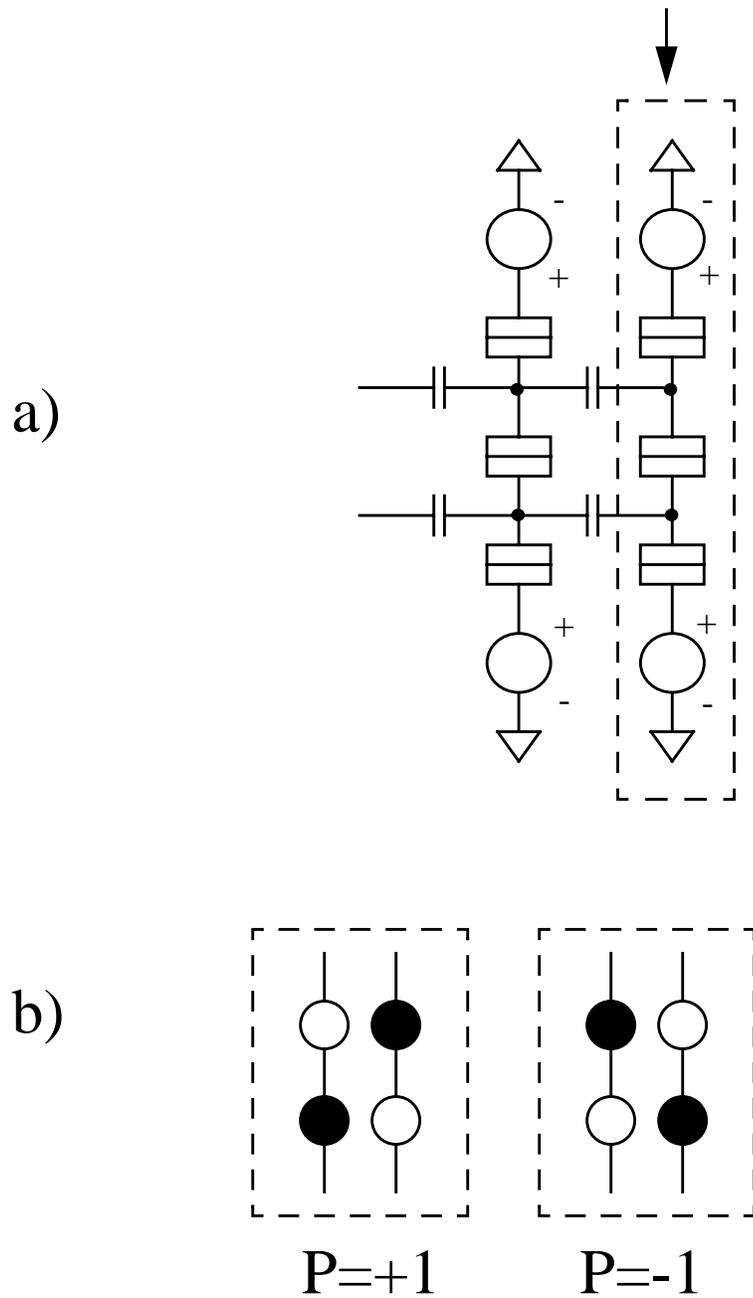

a)

b)

P=+1        P=-1

Figure 6

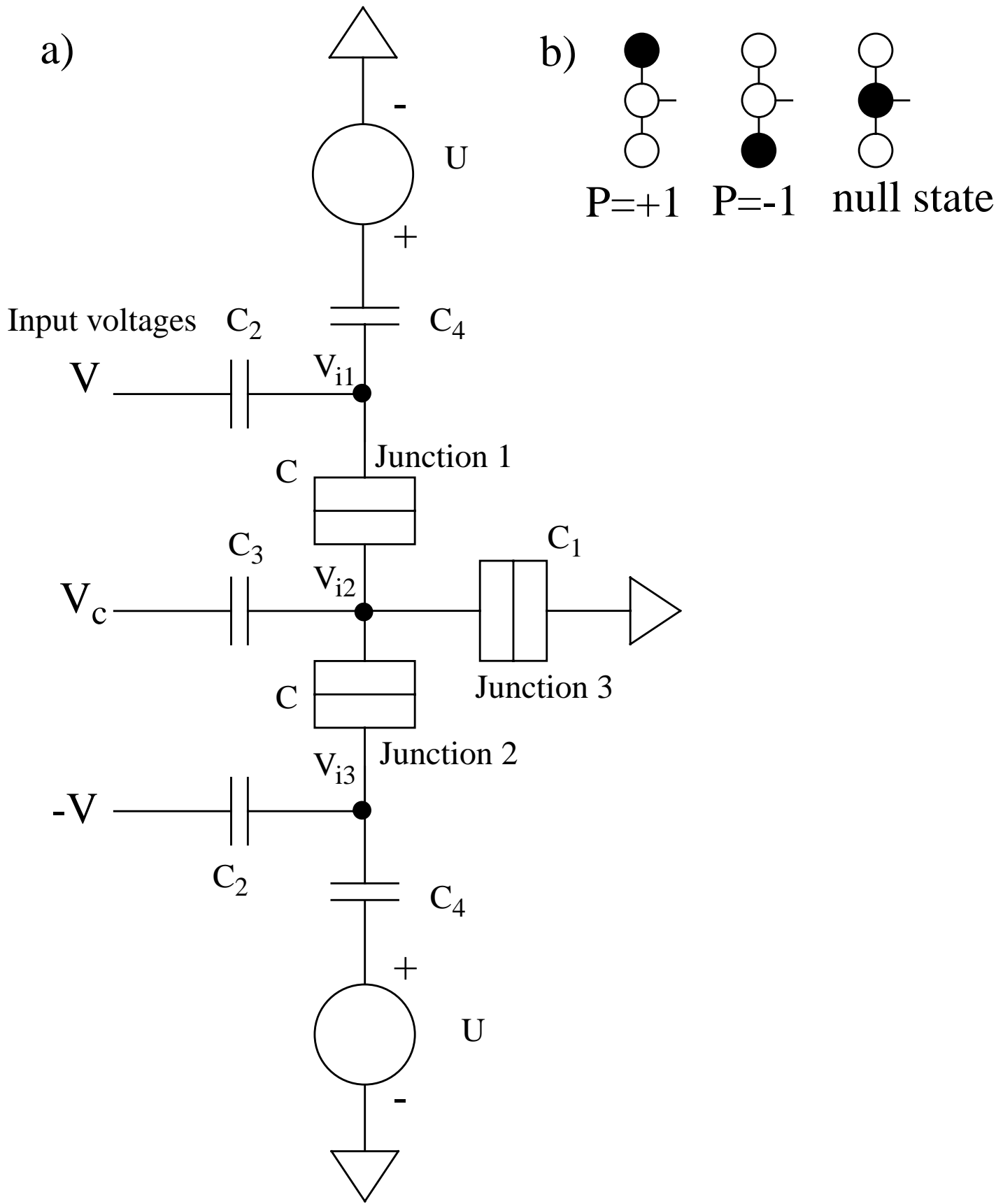

Figure 7

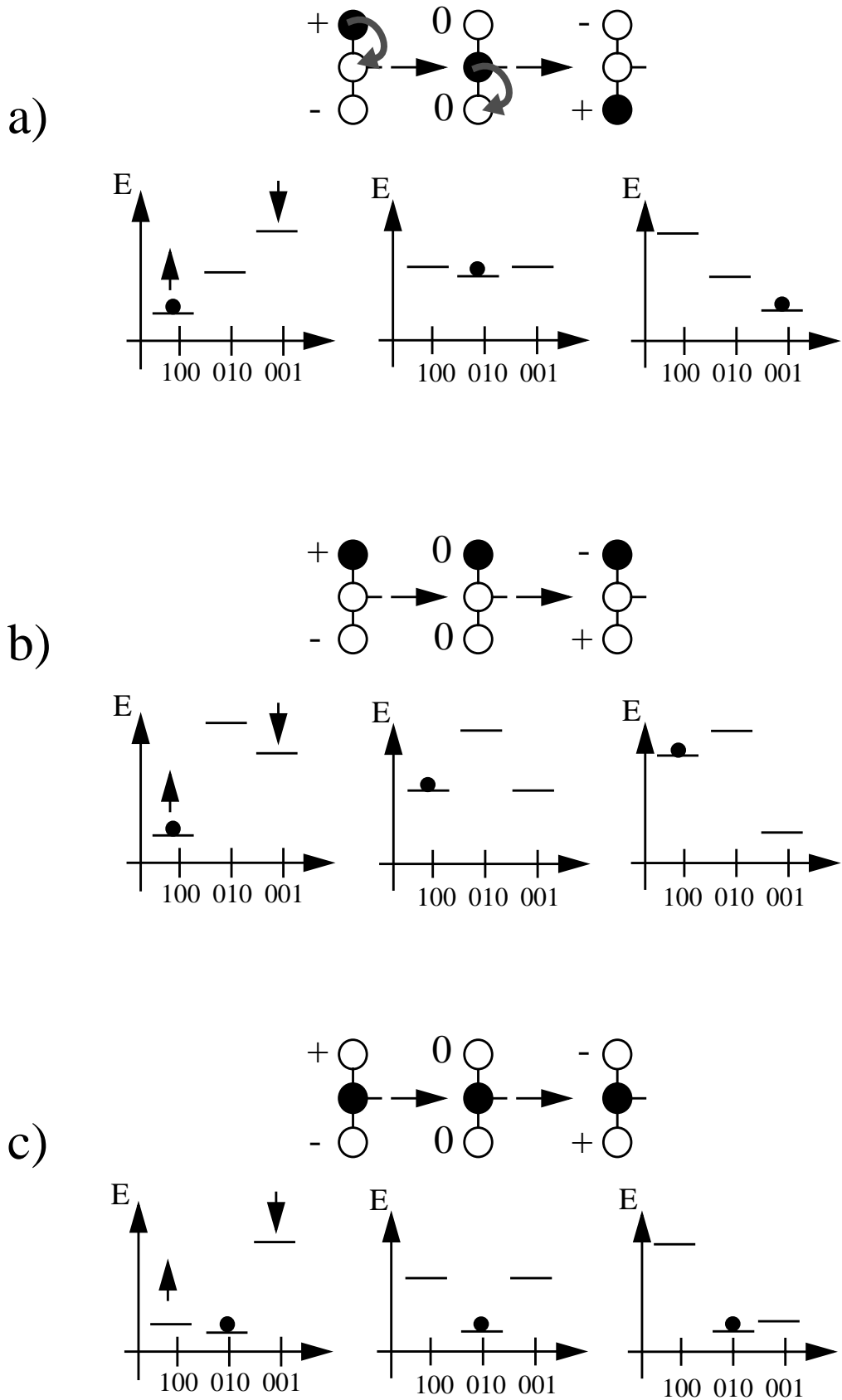

a)

b)

c)

Figure 8

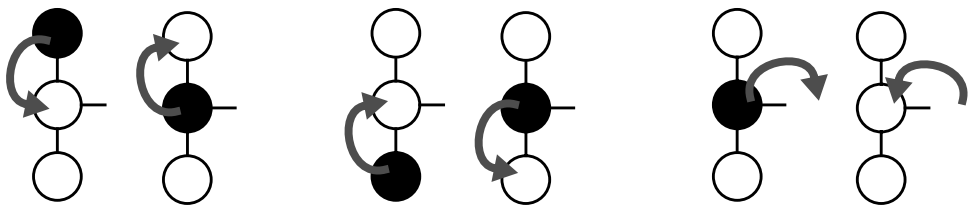

Figure 9

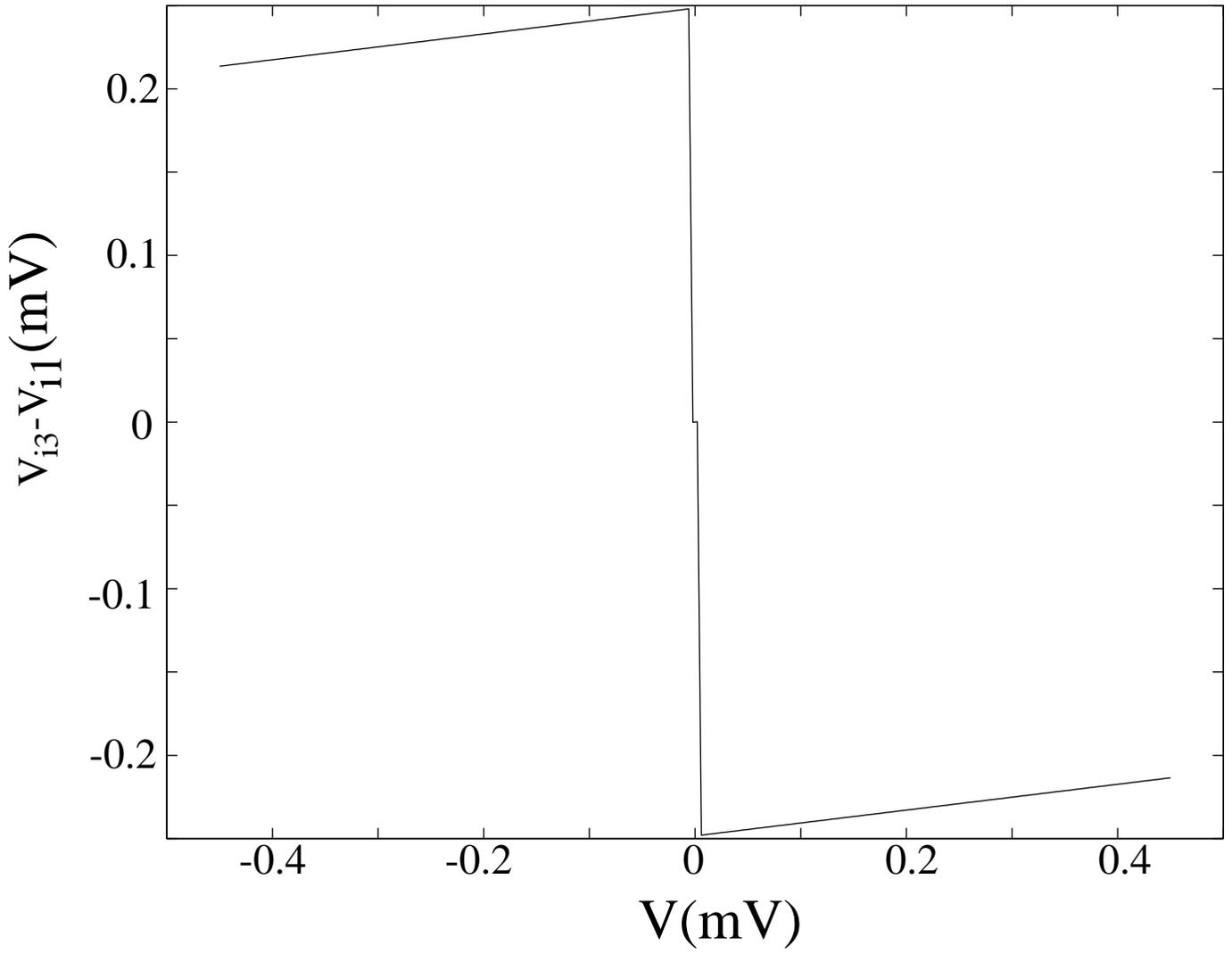

Figure 10

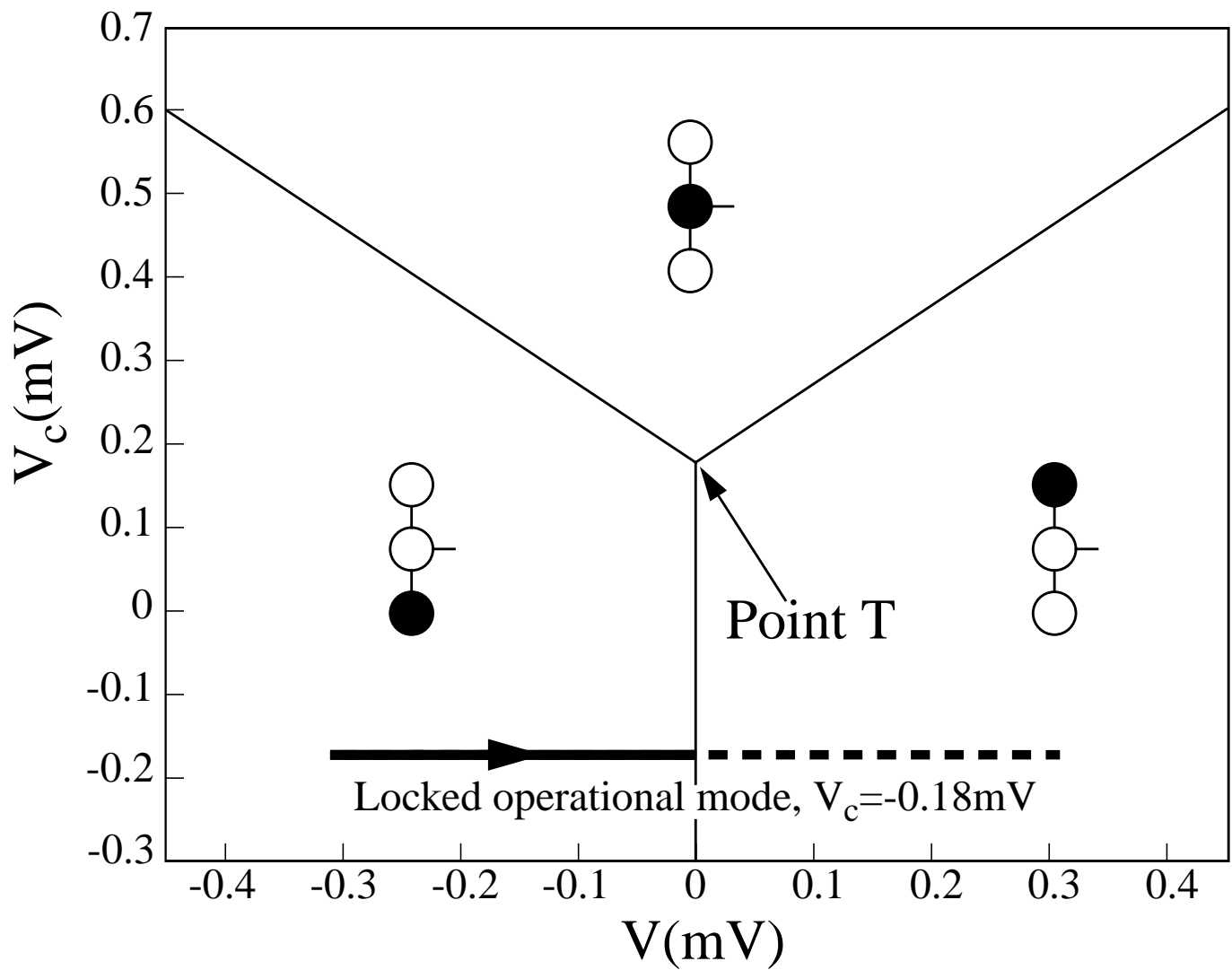

Point T

Locked operational mode, $V_c = -0.18$mV

Figure 11

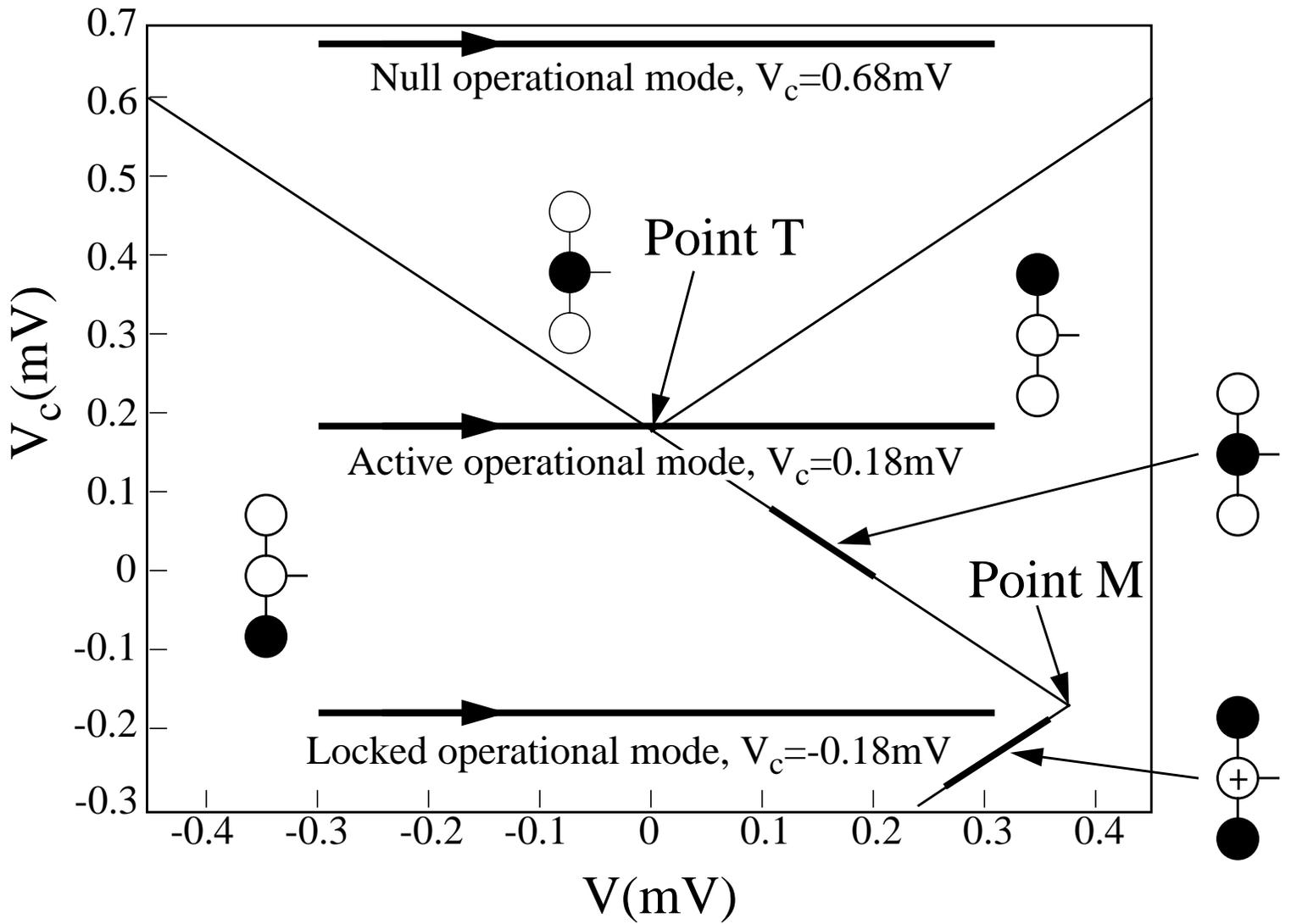

Figure 12

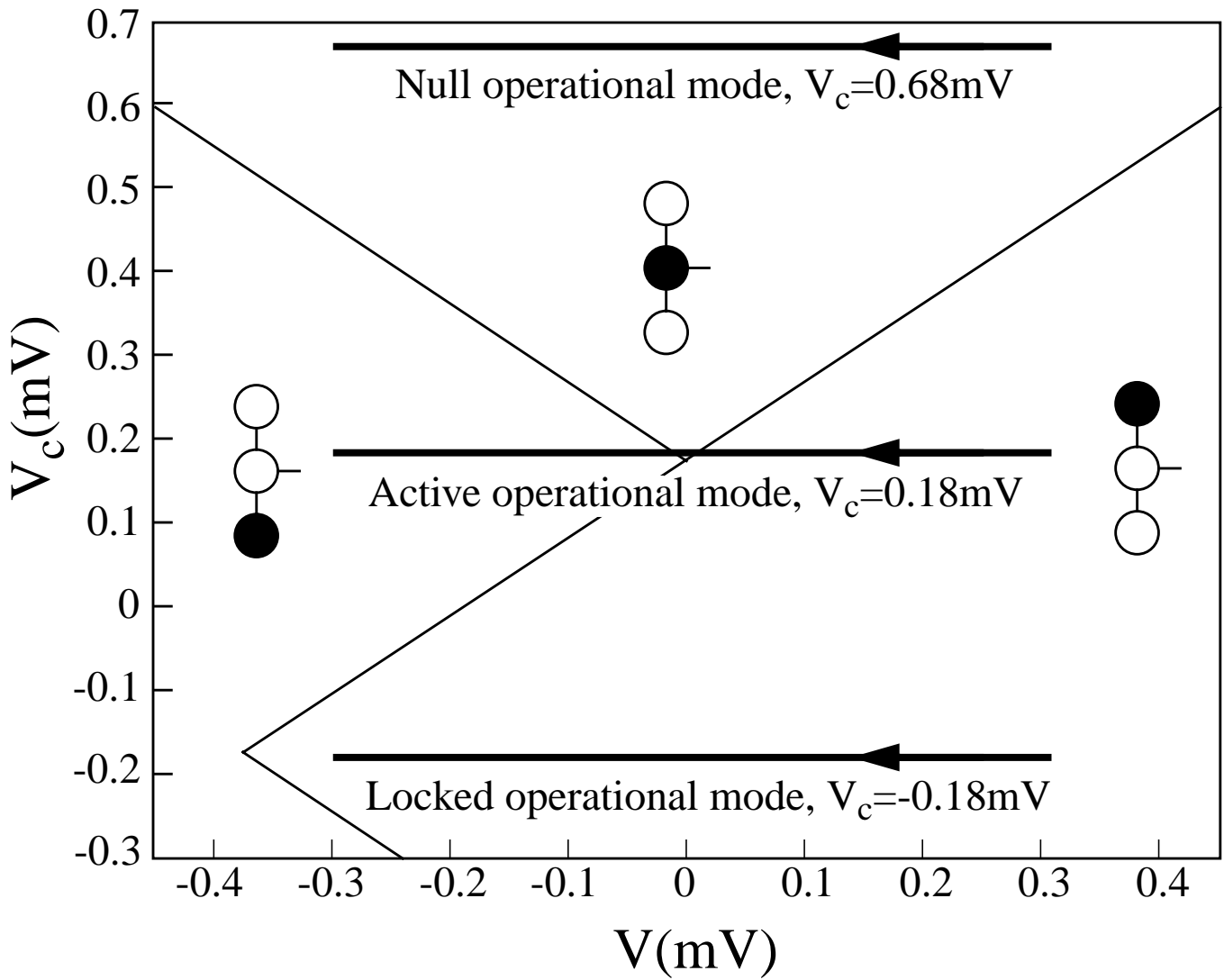

Figure 13

a)

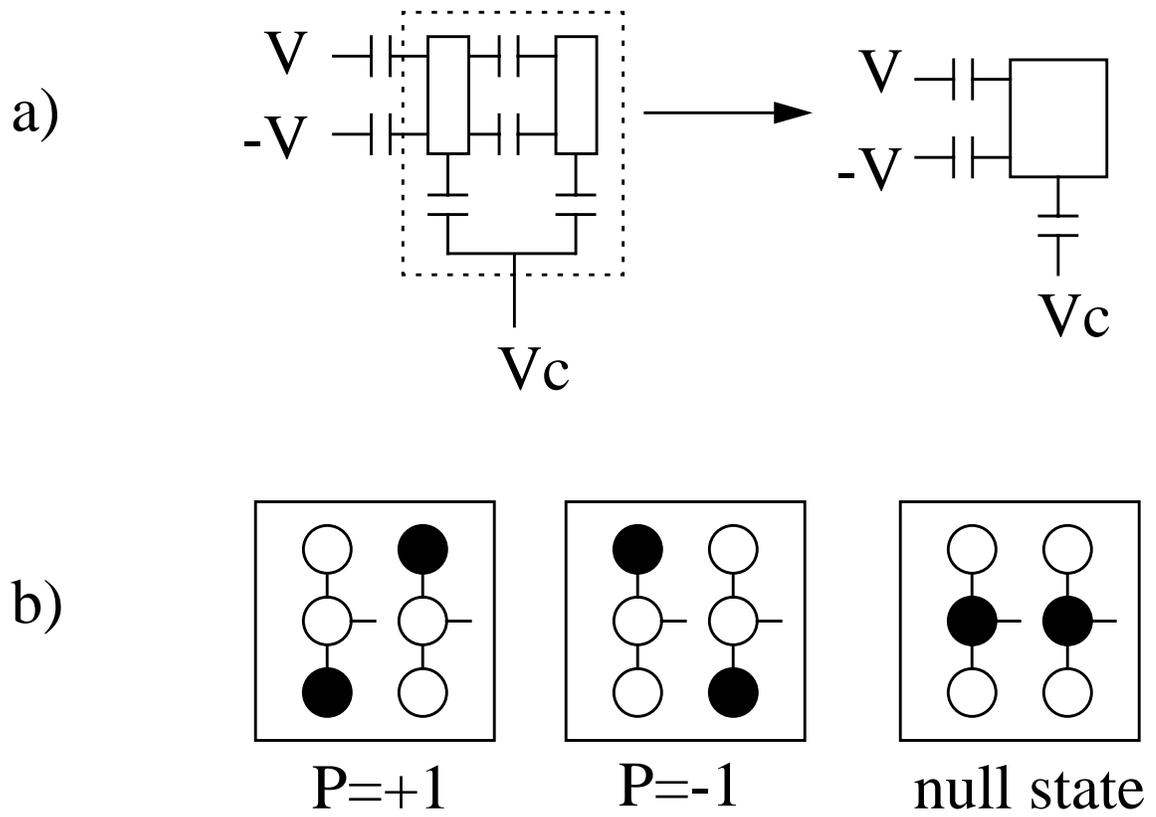

b)

P=+1          P=-1          null state

Figure 14

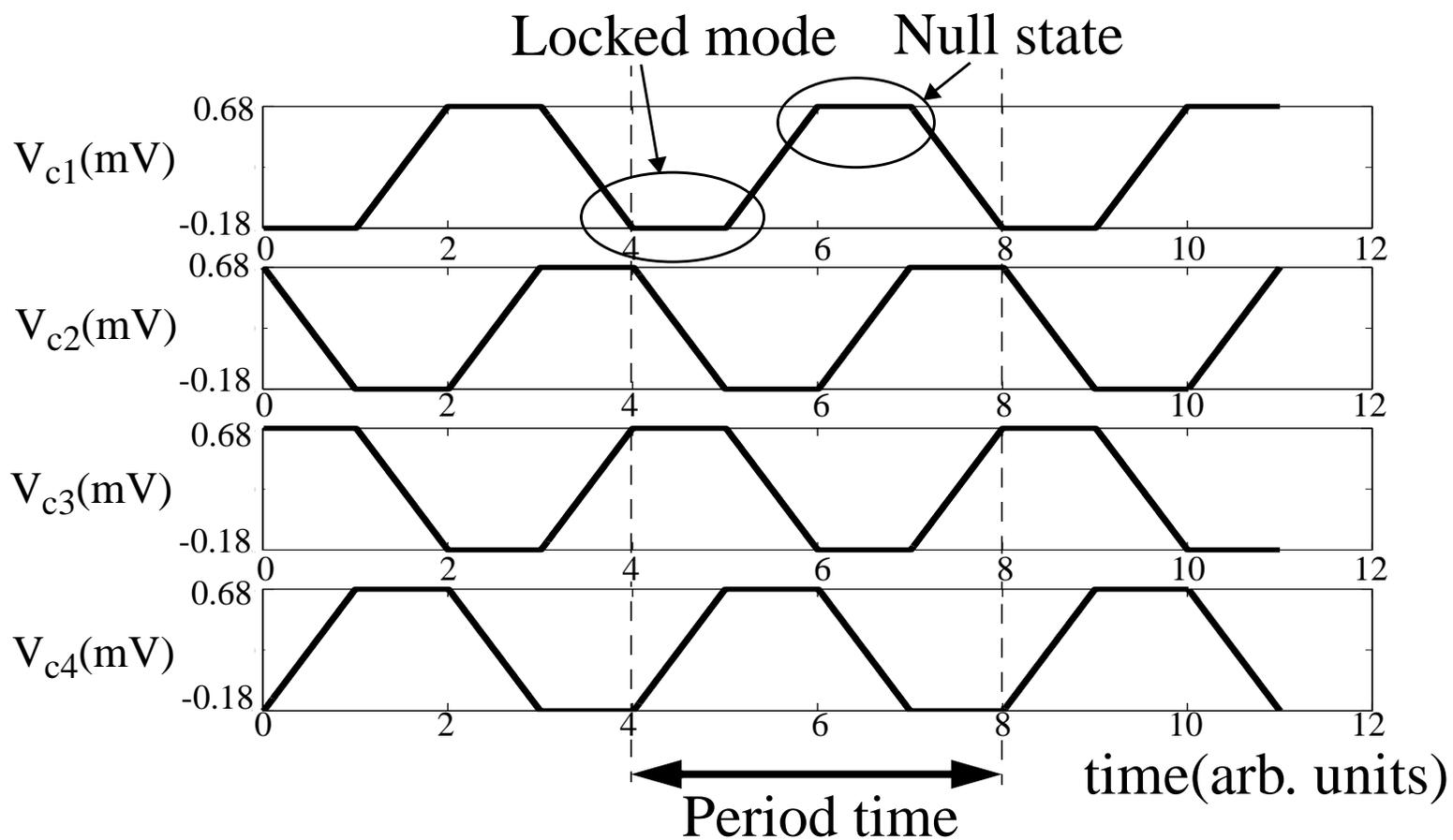

a)

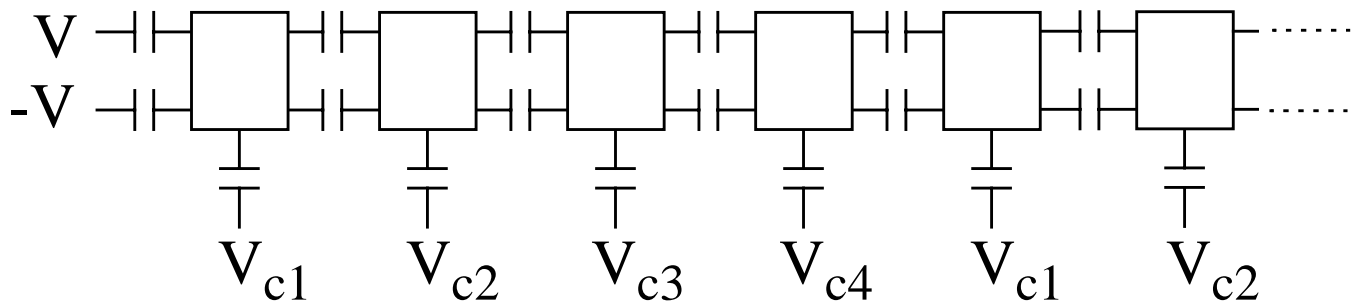

b)

Figure 15

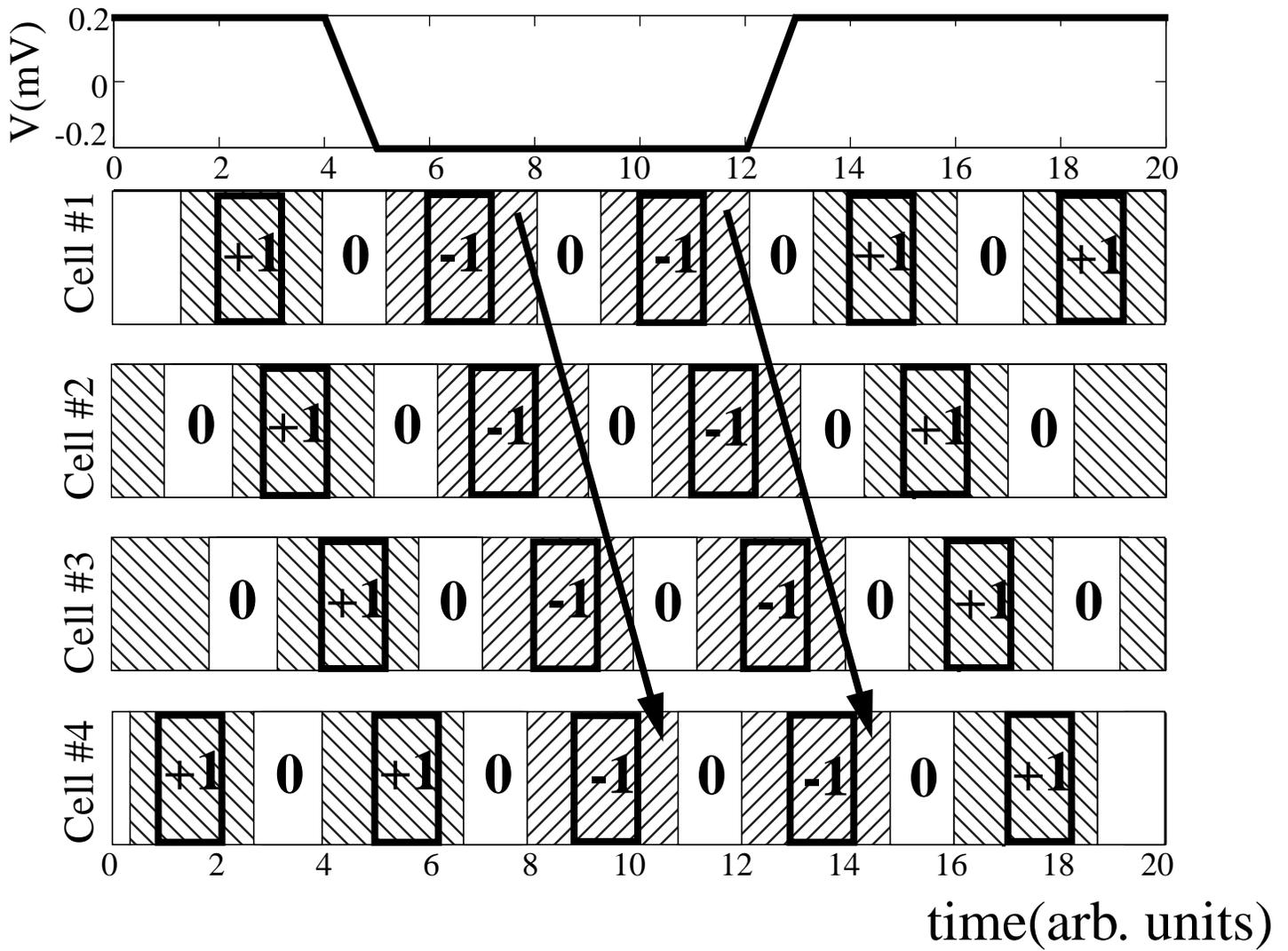

Figure 16